\documentclass[submission,copyright,creativecommons]{eptcs}
\providecommand{\event}{NCMA 2025} 

\usepackage{iftex}

\ifpdf
  \usepackage{underscore}         
  \usepackage[T1]{fontenc}        
\else
  \usepackage{breakurl}           
\fi

\usepackage{enumerate}
\usepackage{paralist}
\usepackage{amssymb}
\usepackage{amsmath}
\usepackage{mathrsfs}
\usepackage{mathtools}
\usepackage{paralist}
\usepackage{hyperref}
\usepackage{algorithm}
\usepackage{algorithmic}

\usepackage[sort&compress,numbers,sectionbib]{natbib}

\usepackage{pgf}
\usepackage{tikz}
\usetikzlibrary{arrows}

\newcommand{\where}{\mid} 
\newcommand{\eps}{\varepsilon} 
\renewcommand{\phi}{\varphi} 
\renewcommand{\rho}{\varrho} 
\newcommand{\Ra}{\Rightarrow} 

\DeclareMathOperator{\aalph}{alph}
\DeclareMathOperator{\card}{card}

\newtheorem{definition}{Definition}{\bf}{\rm}

\newcommand{\footnoteref}[1]{\textsuperscript{\ref{#1}}}

\title{Orchestration of Music by Grammar Systems}
\author{Jozef Makiš
\institute{Faculty of Information Technology,\\ Brno University of Technology,\\ Brno, Czech republic}
\email{xmakis00@stud.fit.vut.cz}
\and
Alexander Meduna
\institute{Faculty of Information Technology,\\ Brno University of Technology,\\ Brno, Czech republic}
\email{meduna@fit.vut.cz}
\and
Zbyněk Křivka
\institute{Faculty of Information Technology,\\ Brno University of Technology,\\ Brno, Czech republic}
\email{krivka@fit.vut.cz}
}

\def\titlerunning{Orchestration of Music by Grammar Systems}
\def\authorrunning{J. Makiš, A. Meduna \& Z. Křivka}

\begin{document}



\maketitle

\begin{abstract}
This application-oriented study concerns computational musicology, which makes use of grammar systems. We define multi-generative rule-synchronized scattered-context grammar systems (without erasing rules) and demonstrates how to simultaneously make the arrangement of a musical composition for performance by a whole orchestra, consisting of several instruments. Primarily, an orchestration like this is illustrated by examples in terms of classical music. In addition, the orchestration of jazz compositions is sketched as well. The study concludes its discussion by suggesting five open problem areas related to this way of orchestration.
\end{abstract}


\section{Introduction}
\label{sec:Introduction}
Formal languages and their models, such as automata and grammars, represent a well-developed body of knowledge, which fulfill a crucially important role in theoretical computer science as a whole. Indeed, these models, such as Turing machines, have allowed this science to establish the very fundamentals of computation, including such key areas as computability, decidability, or computational complexity. From a practical viewpoint, there also exist engineering applications of these models; for instance,  compiler writing customarily makes use of finite  and pushdown automata, regular expressions, and context-free grammars. Nevertheless, admittedly, the significance of these models in theory somewhat exceeds that of their use in practice. To reduce this theory-versus-practice imbalance, researchers have struggled to use and apply these models in a variety of creative areas concerning not only science but also art, such as visual art made by automata (see~\cite{adamatzky2016designing}). Recently, researchers have also studied how to use automata or grammars, such as classical generative grammars or L systems, in musicology (see~\cite{lukas2006multigenerative, PankhurstSonataForm, jurish2004music, humphreys2021investigation, albarracin2021using, eibensteiner2018procedural, melkonian2019music, jin2013formal, gilbert2007probabilistic, bel1992modelling, keller2007grammatical, zuidema2018formal, krakowski2009rhythmically, manousakis2006musical, edwards2011algorithmic, worth2005growing, mccormack1996grammar, manousakis2009non, gogins2006score, prusinkiewicz1986score}). The present paper contributes to this modern application-oriented trend concerning the use of language models to compose music.

Up until now, all the studies concerning the use of language models in music have restricted their investigation to the composition of a music score for a single instrument, such as piano. The fundamental goal of the present paper consists in a generalization of this investigation so it simultaneously produce a score for several instruments. In other words, this application-oriented study demonstrates how to make the arrangement of a musical composition for performance by a whole orchestra.  Simply and plainly put, it shows how to orchestrate music based upon language models.
	
More specifically, consider an $n$-instrument orchestra, where $n$ is a natural number; for example, for a nonet, $n$ = 9. In this paper, we describe how to produce a score for this orchestra by using a grammar system consisting of $n$ grammatical components, represented by scattered context grammars (without erasing rules) in this paper.  In terms of the orchestra, every component corresponds to one of the $n$ instruments, and its goal consists in the generation of the score for the corresponding instrument. During a generative step made by the $n$-component system, all the components work in parallel, and the selection of the rules applied in every single component is globally synchronized across the system as a whole. This synchronization is arranged by a finite number of prescribed $n$-rule sequences so that the system selects one of these sequences and applies its $j$th rule in the $i$th component, $1 \le i \le n$. Once a sequence of $n$ terminal strings is generated by repeatedly making generative steps in the way sketched above, the generative process stops. From a musicological standpoint, the resulting sequence generated in this way represents the score for the whole $n$-instrument orchestra in such a way that the $i$th terminal string represents the score for the $i$th instrument.

The present paper is organized as follows. Section \ref{sec:Preliminaries} recalls all the terminology needed in this paper. Section \ref{sec:Definitions} defines the notion of a rule-synchronized grammar system with scattered context components. Section \ref{sec:Orchestration}, which represents the heart of the present study, explains how to use these systems to generate multi-instrument score.  Section \ref{sec:Example} illustrates this by an example. Section \ref{sec:Evaluation} evaluates the proposed method in the context of music generation using formal models. Section \ref{sec:Conclusion} closes all the study by its summarization and a formulation of important open problem areas concerning the subject of this paper.

\section{Preliminaries}
\label{sec:Preliminaries}
We assume that the reader is familiar with discrete mathematics, and formal theory
(see~\cite{AU,MAH}) as well as formal language theory (see~\cite{Me14,FL,HoFL}).

For a set~$W$, $\card(W)$ denotes its \emph{cardinality}.
An \emph{alphabet} is a finite nonempty set---elements are called \emph{symbols}.
Let $V$ be an \emph{alphabet}. $V^*$ is the \emph{set of all strings} over $V$.
Algebraically, $V^*$ represents the free monoid generated by~$V$ under the operation
of concatenation. The identity of~$V^*$ is denoted by~$\eps$.
Set $V^+ = V^* - \{\eps\}$. Algebraically, $V^+$ is thus the free semigroup
generated by~$V$ under the operation of concatenation.
For~$w \in V^*$, $a \in V$, and $A \subseteq V$, $|w|$ denotes the \emph{length of~$w$},
$\#_a(w)$ denotes the \emph{number of occurrences of the symbol $a$ in $w$},
and $\#_A(w)$ denotes the \emph{number of occurrences of the symbols from $A$ in $w$}.
The \emph{alphabet of $w$}, denoted by $\aalph(w)$, is the set of symbols appearing in~$w$.

Let $\Ra$ be a relation over $V^*$. We denote $i$th power of $\Ra$ as $\Ra^i$, for $i \geq 0$. The transitive and the transitive-reflexive closure of $\Ra$ are denoted by $\Ra^+$ and $\Ra^*$, respectively. Unless we explicitly stated otherwise, we write $x\ \Ra\ y$ instead of $(x,y) \in\ \Ra$ throughout.

\section{Definitions}
\label{sec:Definitions}
The present section defines the language theory notions used throughout the rest of this paper. First, it defines scattered context grammars, which represent well-known grammatical model. Then, based upon these grammars, it introduces rule-synchronized music grammar systems, which are later used as an orchestration formalism for music.
\begin{definition}
\label{def:scg}
A \emph{scattered context grammar} is a quadruple, $G = (N, T, P, S)$, where $N$ and $T$ are alphabets such that $N\cap T =\emptyset$.  Symbols in $N$ are referred to as nonterminals while symbols in $T$ are terminals.  $N$ contains $S$---the start symbol of $G$.  $P$ is a finite non-empty set of rules such that every $p \in P$ has the form $$(A_1,\allowbreak\dots,A_n)\rightarrow(x_1,\allowbreak\dots,x_n),$$ where $n  \geq 1$, and for all $i = 1, \allowbreak\dots, n$, $A_i \in N$ and $x_i  \in(N\cup T)^*$. If each $x_i$ satisfies $|x_i| \leq 1$, $i = 1,\allowbreak\dots, n$,  then  $( A_1, \allowbreak\dots, A_n) \rightarrow  (x_1, \allowbreak\dots, x_n)$ is said to be simple. If $n = 1$, then $(A_1) \rightarrow  (x_1)$ is referred to as a context-free rule; for brevity, we hereafter write $A_1 \rightarrow  x_1$ instead of $(A_1) \rightarrow  (x_1)$.  If for some $n \geq 1$, $(A_1, \allowbreak\dots, A_n) \rightarrow  ( x_1, \allowbreak\dots, x_n) \in P$, $v = u_1A_1u_2\allowbreak\cdots u_{n-1}A_nu_n$, and $w = u_1x_1u_2\allowbreak\cdots u_{n-1}x_nu$ with $u_i \in(N  \cup T)^*$ for all $i = 1, \allowbreak\dots, n$, then $v$ directly derives $w$ in $G$, symbolically written as $v \Rightarrow w\ [(A_1, \allowbreak\dots, A_n) \rightarrow (x_1, \allowbreak\dots, x_n)]$ or, simply, $v \Rightarrow w$ in $G$. In the standard manner, extend $\Rightarrow$ to $\Rightarrow^n$, where $n \geq 0$; then, based on $\Rightarrow^n$, define $\Rightarrow^+$  and $\Rightarrow^*$.
The language of $G$, $L(G)$, is defined as $L(G) = \{w  \in T^*\where  S  \Rightarrow^* w\}$.  A derivation of the form $S  \Rightarrow^* w$ with $w \in T^*$ is called a successful derivation.
\end{definition}

Next, we define the notion of an $n$-generative rule-synchronized music grammar system as the central notion of this paper as a whole. In essence, this notion is based upon that of an $n$-generative rule-synchronized music grammar system with context-free components (see \cite{lukas2006multigenerative} and Section 13.3 in \cite{meduna2020handbook}), but the new notion is underlain by scattered context components.

\begin{definition}
\label{def:ngrsg}
An $n$-generative rule-synchronized music grammar system is defined as an $(m+1)$-tuple $$G_s = (G_1, \ldots, G_m, Q),$$ in which

\begin{itemize}
    \item{$G_i = (N_i, T_i, P_i, S_i)$ is a scattered context grammar introduced in Definition \ref{def:scg}, for all $i = 1, \ldots, m$;}
    \item{$Q$ is a finite set that consists of n-tuples structured as $(p_1, p_2, \ldots, p_m)$, where $p_i \in P_i$, for all $i = 1, \ldots, m$.}
\end{itemize}
\end{definition}

In addition to the original definition, we will use tokens instead of plain terminals. Tokens have indexed attributes they represent that are going to be taken into account in the final music interpretation by the instrument. Tokens are in the form $t_{[w_1, w_2, \ldots, w_k]} \in T_i$, where $w_1, w_2, \ldots, w_k$ are music attributes like tone length, special operation (tone inversion, shift, etc.), chord or others. Number~$k$ expresses the number of token attributes.

To improve readability while generating harmonic passages in music, we chose to represent chords using symbols from the Greek alphabet for simplicity, as they are difficult to denote with single-character symbols. In the example, there are mappings of symbols from Greek alphabet to chords.

The terminal strings derived from the start symbol of a grammar or in our model are in $m$-form as $m$-tuples structured as $S_f = (x_1, \ldots, x_m)$, where $x_i \in T^*$, for all $i = 1, \ldots, m$. Let us take
$$c_i = a_1A_1\cdots a_{n-1}A_na_{n},$$
$$d_i = a_1x_1\cdots a_{n-1}x_na_{n}.$$ Then $S_f = (c_1, c_2, \ldots, c_m)$ and $\bar{S}_f = (d_1, d_2, \ldots, d_m)$ are sentential $m$-forms, in which $c_i,d_i \in (N \cup T)^*$, for every $i = 1, \ldots, m$. Consider $r_i$: $(A_1,\ldots,A_n) \rightarrow (x_1,\ldots,x_n)$ $\in P_i$ for all $i = 1, \ldots, m$ and $(r_1, r_2, \ldots, r_m)$ $\in Q$, such that $r_i = c_i \rightarrow d_i$. Consequently, $S_f$ directly derives $\bar{S}_f$ in $G_s$, denoted by $$S_f \Rightarrow_{G_s} \bar{S}_f.$$

Let us generalize $\Rightarrow_{G_s}$ with $\Rightarrow_{G_s}^k$, for all $k \geq 0$, $\Rightarrow_{G_s}^+$ and $\Rightarrow_{G_s}^*$. Generated $m$-string of $G_s$, denoted by \textnormal{m-}$S(G_s)$, we define by $$m\textnormal{-S}(G_s) = \{(w_1, \ldots, w_m) \where (S_1, \ldots, S_m) \Rightarrow_{G_s}^* (w_1, \ldots, w_m),$$ $$w_i \in T^*, \textnormal{for all} \, i = 1, \ldots, m\}.$$

\section{Orchestration}
\label{sec:Orchestration}
Building on the concepts and formalisms introduced in the previous section, this part of the work is focused on the orchestration process across multiple instruments. What led us to this is the work of others that are dealing with the algorithmic composition and grammar-based music generation. The popularity of grammar-based approaches has started with interesting applications using L-systems \cite{prusinkiewicz1986score} where generated string is interpreted as a sequence of notes. This research was expanded in the works of \cite{gogins2006score,rodrigues2016evolving,mccormack1996grammar,worth2005growing,edwards2011algorithmic} and many others. A doctoral dissertation explored automata driven by rhythm in musical improvisation \cite{krakowski2009rhythmically}. It may seem like the L-systems rule the grammar-based approaches but that is just not true. The diversity of grammatical frameworks has been explored in the literature. For instance, \cite{zuidema2018formal} investigates hierarchical structure-building mechanisms across music, language, and animal song using formal language theory. By using context-free grammars, \cite{keller2007grammatical} describes how to model jazz improvisation within a controlled generative system. The notion of a probabilistic context-free grammar specifically tailored for melodic reduction is discussed in \cite{gilbert2007probabilistic}. Furthermore, \cite{jin2013formal} presents a formal semantic framework to model control flow in Western music notation. Similarly, \cite{melkonian2019music} applies probabilistic temporal graph grammars to model music as a language. In \cite{eibensteiner2018procedural}, a  procedural music generation by using formal grammars is explored. Finally, \cite{humphreys2021investigation} applies grammar-based compression techniques to uncover structural patterns in music.

\begin{figure}[H]
    \centering
    \includegraphics[scale=0.4]{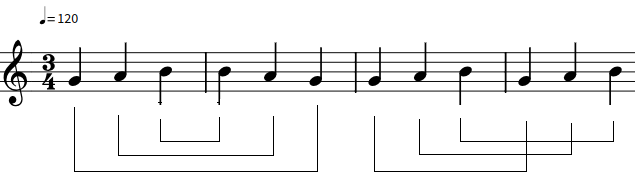}
    \caption{Context-free ($1^{\textnormal{st}}$ half) and non-context free dependencies ($2^{\textnormal{nd}}$ half).}
    \label{fig:mixdepen}
\end{figure}

While some of the cited works are capable of capturing both context-free and non-context-free dependencies (see Fig.~\ref{fig:mixdepen}), as discussed in \cite{jurish2004music}, they fall short when it comes to modeling the complex interactions present in multi-instrumental compositions. By context-free and non-context-free dependencies, we refer to nested and crossing connections between notes, respectively. For this reason, we have chosen to use an $n$-generative rule-synchronized music grammar system, which allows the system to make the simultaneous rewriting of multiple nonterminals. This property makes them well-suited to represent interdependent musical structures that occur in music. As a basic example, we can take the piano, which can have written harmony in the bass clef and written melody in the treble clef. Or two instruments like the piano and violin may complement one another to produce a richer and more engaging melodic texture.

As a component of our grammar system, context-free grammar would not be just enough. As a demonstration, we can take a look at Fig.~\ref{fig:mixdepen}. Starting from context-free grammars, we can describe well-connected melodies. Well-connected melodies go somewhere and return in a similar way, but such structures are not common in music. More commonly, repetition and variation create crossing dependencies, such as the ones we can see in the second half of the figure. This approach fits classical and jazz music, but it can be applied almost in any structural music.

\subsection*{Encoding Musical Concepts into the Grammar}
\label{subsec:encodingintorules}
To showcase our model, we have picked the sonata form from classical music, and jazz music is represented by its standard form. Mentioned forms presented here are taken from \cite{ravelliJazzForm2025} and \cite{Kadlec2022}.

We have decided to talk about two examples to demonstrate how musical pieces could be encoded into grammar. The first is popular jazz song Take The A Train from \cite{ravelliJazzForm2025}. The second is \cite{PankhurstSonataForm} and shows a minimalistic example of sonata form called Allegro in F composed by Mozart.

When choosing a top-down approach to analyze a musical piece, we start by examining its overall structure. A great example is the jazz song \cite{ravelliJazzForm2025}, which uses the most common structure in jazz standards, the $AABA$ form. This song consists of two distinct sections ($A$ and $B$), with each section typically spanning eight measures. These sections form the standard 32-measure framework of the basic melody found in $AABA$ jazz compositions. 

When applying a similar analytical approach to the sonata form, we observe a three-part structure: exposition ($A$), development ($B$), and recapitulation ($A'$). The exposition introduces the primary thematic material, typically divided into two contrasting themes. The development explores these themes through variations, modulations, and transformations. Finally, the recapitulation returns to the original thematic material, usually restating the exposition themes in their original keys or slightly modified. This structured approach allows composers to achieve a coherent and varied musical narrative, which is fundamental to classical sonata compositions.

The from can vary in different compositions, styles. For example, we can generate the $AABA$ or $ABA'$ form with following rules:
\begin{align*}
S &\rightarrow AABA \\
\text{or} \quad S &\rightarrow ABA'.
\end{align*}

\subsection*{Encoding Melody and Harmony}
Once we have generated the initial nonterminals that outline the structure of the musical piece, the next step is to create the actual musical content. Music is truly creative, and there are endless possibilities. In our sonata example, we could encode exposition into three non-terminals $T_1, R, T_2$ and similarly recapitulation $T'_1, R', T'_2$. The symbol $R$ represents the transitions between the tonic and dominant phrases $T_1$ and $T_2$. $T_1$ and $T_2$ are also themes of our song that create interesting tension. Development in an example could be characterized by two variations of original theme and we will denote it by $V_1$ and $V_2$. To put this into rules 
\begin{align*}
(A,A') &\rightarrow (T_1RT_2,\; T'_1R'T'_2), \\
\quad B &\rightarrow (V_1,\; V_2).
\end{align*}

For our jazz example, we first introduce the main theme and then repeat it, perhaps with slight variations. These two $A$ sections are followed by a section known as the bridge, characterized by contrasting melody or harmony. Finally, the original main theme returns. Each of these sections typically consists of eight measures. In the jazz piece we have selected we have a theme from two similar melodies. Rules that would generate structure would look like:
\begin{align*}
(A,A,A) &\rightarrow (T_1T_2,\; T_1T_2,\; T_1T_2), \\
\quad B &\rightarrow (V_1,\; V_2).
\end{align*}

The last missing piece of a grammar that could generate our example is to define notes to be played in mentioned melodic sections. Sonata rules for the first two measures would look like 
\begin{align*}
(T_1, T'_1) &\rightarrow (d_{[e,2]} h_{[e,1]} a_{[e,2]} c_{[e,2]},\; d_{[e,2]}h_{[e,1]}a_{[e,1]} c_{[e,2]}), \\[6pt]
(T_1, T'_1) &\rightarrow (h_{[e,1]}a_{[e,-1]}p_{[e,-1]}c_{[e,1]},\; h_{[e,1]}a_{[e,-2]}p_{[e,-1]}c_{[e,2]}).
\end{align*}

On the right-hand side of the grammar rules, tone names are indexed using brackets, where the first symbol (e) indicates note duration (length—in this case, an eighth note), and the second number specifies the pitch interval or position within the current musical context.

For simplicity this model, is not meant to analyze the musical structure beyond the level of a single measure. This approach helps to ensure rhythmic consistency in the generated music and provides a clearer, more polished grammatical representation. Additionally, it eliminates the need to calculate the exact number of beats per measure or manage the filling of any remaining rhythmic gaps. The presented approach could be applied to any musical piece. We define our form, and after that, from form, we can generate various numbers of melodic and harmonic passages. Formally, this can be represented by grammar rules of the following general structure:

\begin{align*}
(A,A') &\rightarrow (T_1H_1T_2H_2,\; T'_1H'_1T'_2H'_2), \\
\quad B &\rightarrow (V_1H_1,\; V_2H_2).
\end{align*}

Here, we have a characterization of a musical piece that features a switch between tonic and harmonic sections. Followed by different variations that could be picked up from classical composers like Bach, Beethoven and others. This is a creative process, and it is up to the creator of the grammar to determine how their music is perceived.

\subsection*{Encoding Multi-Instrumental Compositions into Grammar Rules}
\begin{figure}[H]
    \centering
    \includegraphics[scale=0.3]{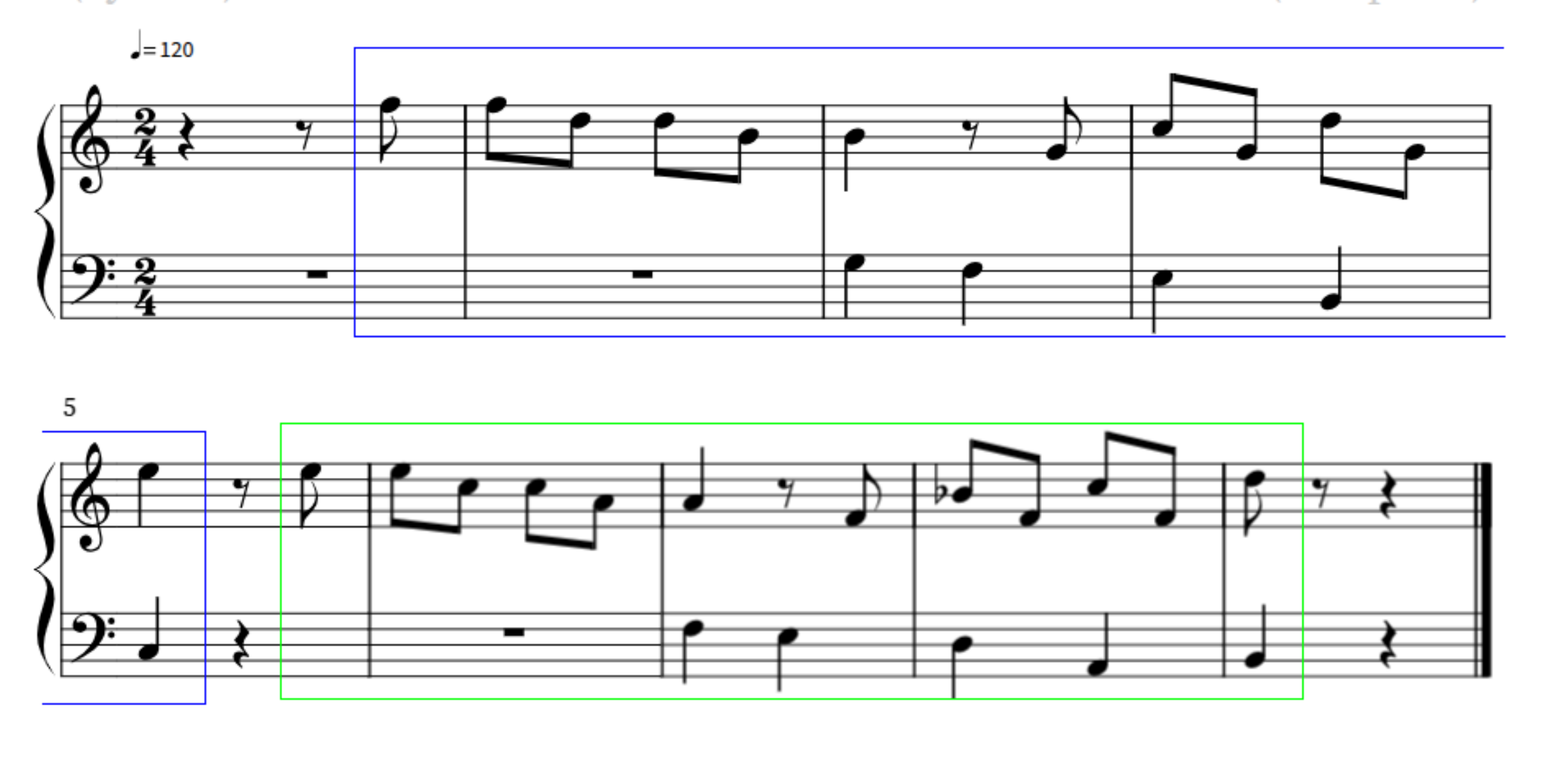}
    \caption{A small example of dependencies between music staffs.}
    \label{fig:sonatamultistring}
\end{figure}

We have covered how to create a musical piece when there is only one instrument and needs only one staff. For example, a piano has two staffs. Of course, a staff can still be interpreted by an instrument, but it would lack melody or harmony. From Figure \ref{fig:sonatamultistring}, we can see how important it is to have a model that is able to synchronize the generation of music between treble and bass clefs for piano. The bass clef mirrors the melody created by treble clef. For this reason, we use a rule-synchronized model that ensures these properties are preserved. A similar approach can be applied to music for multiple instruments, where instruments often copy the melody, create contrast, create tension, or use other musical expressions to make music interesting.

Figure \ref{fig:sonatamultistring} comes from the development of \cite{PankhurstSonataForm}. The first rectangle (green) is a variation of the notes selected in the second rectangle (blue). This can be easily encoded into a 2-component system:
$$G_s = (G_1, G_2, Q),$$ where

\begin{itemize}
    \item{$G_1 = (\{S_1, T, T_{\downarrow}\},\, \{r_{[-, q, -]}, r_{[-, e, -]}, f_{[-, e, 2]}, d_{[-, e, 2]}, h_{[-, e, 1]}, f_{[\downarrow, e, 2]}, \\ d_{[\downarrow, e, 2]}, h_{[\downarrow, e, 1]}, h_{[-, q, 1]}, g_{[-, e, 1]}, c_{[-, e, 2]}, c_{[\downarrow, e, 2]}, g_{[\downarrow, e, 1]}, h_{[\downarrow, q, 1]}, e_{[-, q, 2]}, e_{[-, e, 2]}, \\ r_{[-, q, -]}\},\, 
    \\ \{1\colon S_1 \rightarrow (r_{[-, q, -]}r_{[-, e, -]}f_{[-, e, -]}T, T_{\downarrow}),  
    \\ 2\colon (T, T_{\downarrow}) \rightarrow (f_{[-, e, 2]}d_{[-, e, 2]}d_{[-, e, 2]}h_{[-, e, 1]}T, f_{[\downarrow, e, 2]}d_{[\downarrow, e, 2]}d_{[\downarrow, e, 2]}h_{[\downarrow, e, 1]}T_{\downarrow}), 
    \\ 3\colon (T, T_{\downarrow}) \rightarrow (h_{[-, q, 1]}r_{[-, e, -]}g_{[-, e, 1]}T, h_{[\downarrow, q, 1]}r_{[-, e, -]}g_{[\downarrow, e, 1]}T_{\downarrow}), 
    \\ 4\colon (T, T_{\downarrow}) \rightarrow (c_{[-, e, 2]}g_{[-, e, 1]}d_{[-, e, 2]}g_{[-, e, 1]}T, c_{[\downarrow, e, 2]}g_{[\downarrow, e, 1]}d_{[\downarrow, e, 2]}g_{[\downarrow, e, 1]}T_{\downarrow}),
    \\ 5\colon (T, T_{\downarrow}) \rightarrow (e_{[-, q, 2]}r_{[-, e, -]}e_{[-, e, 2]}, e_{[-, e, 2]}r_{[-, e, -]}r_{[-, q, -]})
    \}, S_1)$}
    \item{$G_2 = (\{S_2, B, B_{\downarrow}\},\, \{r_{[-, h, -]}, g_{[-, q, -]}, f_{[-, q, -]}, g_{[\downarrow, q, -]}, \\ f_{[\downarrow, q, -]}, e_{[-, q, -]}, h_{[-, q, -]}, e_{[\downarrow, q, -]}, h_{[\downarrow, q, -]}, a_{[-, q, -]}, r_{[-, q, -]}, a_{[\downarrow, q, -]}, r_{[-, q, -]}\}, \,
    \\\{1\colon S_2 \rightarrow (r_{[-, h, -]}B, B_{\downarrow}),
    \\ 2\colon (B, B_{\downarrow}) \rightarrow (r_{[-, h, -]}B, r_{[-, h, -]}B_{\downarrow})\}
    \\ 3\colon (B, B_{\downarrow}) \rightarrow (g_{[-, q, -]}f_{[-, q, -]}B, g_{[\downarrow, q, -]}f_{[\downarrow, q, -]}B_{\downarrow}) 
    \\ 4\colon (B, B_{\downarrow}) \rightarrow (e_{[-, q, -]}h_{[-, q, -]}B, e_{[\downarrow, q, -]}h_{[\downarrow, q, -]}B_{\downarrow}), 
    \\ 5\colon (B, B_{\downarrow}) \rightarrow (a_{[-, q, -]}r_{[-, q, -]}, a_{[\downarrow, q, -]}r_{[-, q, -]})\}, S_2)$}
    \item{$Q = \{(1, 1), (2, 2), (3, 3), (4, 4), (5, 5)\}$.}
\end{itemize}

This shows how easy it is to encode one of the most popular classical songs into the grammar. Grammar $G_1$ has rules that can be applied to generate the treble clef for piano and $G_2$ produces the bass clef. Each measure for both treble and bass clefs is synchronized in the set $Q$.

\subsection*{Derivation Process in Multi-Generative Grammar}
\label{subsec:derivproc}
With the intention to create a music piece, rules have to be applied in a certain order. First, we rewrite starting symbol with nonterminals to define structure of the composition. With that, we can start to rewrite structure symbols so that final melodies and harmonies can take the form. 

To illustrate this, let us begin with an example that generates jazz music for piano using both the treble and bass clef:
$$G_s = (G_1, G_2, Q),$$ in which 
\begin{itemize}
    \item{$G_1 = (\{S_1, A, B\},\, \{c_{y}, a_{x}, g_{x}, e_{y},f_{x},\alpha_{z},\beta_{z},\gamma_{z},\delta_{z},\epsilon_{w},\zeta_{w}\},\, 
    \\ \{1\colon S_1 \rightarrow (AABAS_1), 2\colon S_1 \rightarrow (AABA), 3\colon (A, A, A) \rightarrow (MH, MH, MH), 
    \\ 4\colon (M, M, M) \rightarrow (c_{y}c_{y}a_{x}g_{x}, c_{y}c_{y}a_{x}g_{x}, c_{y}c_{y}a_{x}g_{x}),
    \\ 5\colon (H, H, H) \rightarrow (\alpha_{z}\beta_{z}\gamma_{z}\delta_{z}, \alpha_{z}\beta_{z}\gamma_{z}\delta_{z}, \alpha_{z}\beta_{z}\gamma_{z}\delta_{z}),
    \\ 6\colon B \rightarrow (M_1H_1), 7\colon (M_1, H_1) \rightarrow (\epsilon_{w}\zeta_{w}, e_{y}a_{x}f_{x}a_{x})
    \}, S_1)$,}
    \item{$G_2 = (\{S_2, A, B, P, L\},\, \{c_{v}, g_{v}, r_{v}, a_{u}, e_{v}, r_{t}\}, \,
    \\\{1\colon S_2 \rightarrow (AABAS_2), 2\colon S_2 \rightarrow (AABA), 3\colon (A, A, A) \rightarrow (PL, PL, PL), 
    \\ 4\colon (P, P, P) \rightarrow (c_{v}g_{v}c_{v}g_{v}r_{v}g_{v}c_{v}g_{v}, c_{v}g_{v}c_{v}g_{v}r_{v}g_{v}c_{v}g_{v}, c_{v}g_{v}c_{v}g_{v}r_{v}g_{v}c_{v}g_{v}),
    \\ 5\colon (L, L, L) \rightarrow (c_{v}a_{u}e_{s}r_{v}r_{t}e_{s}a_{u}, c_{v}a_{u}e_{s}r_{v}r_{t}e_{s}a_{u}, c_{v}a_{u}e_{s}r_{v}r_{t}e_{s}a_{u}),
    \\ 6\colon B \rightarrow (PL), 7\colon (P, L) \rightarrow (c_{v}g_{v}c_{v}g_{v}r_{v}g_{v}c_{v}g_{v}, c_{v}a_{u}e_{s}r_{v}r_{t}e_{s}a_{u})
    \}, S_2)$,}
    \item{$Q = \{(1, 1), (2, 2), (3, 3), (4, 4), (5, 5), (6, 6), (7, 7)\}$.}
\end{itemize}

Grammars in system $G_{s}$ use substitution of token symbols for better readability in defined grammar and in following derivations. Explanation of the tokens in $G_1$ is in the tables Tab. \ref{Tb1} and Tab. $\ref{Tb2}$.

These tables explain the symbols used in $G_s$ in this section. The first position, for example in $c_{-, q, 2}$ is used for a variation technique that moves the tone. The symbol $q$ represents a quarter note, $e$ an eighth note, and $h$ a half note. The last element specifies the octave in which the note is placed.
\begin{table}[H]
\centering
\begin{minipage}{0.48\textwidth}
\centering
\begin{tabular}{c|c}
Symbol & Note or Chord \\
\hline
$c_y$ & $c_{[-, q, 2]}$ \\
$a_x$ & $a_{[-, q, 1]}$ \\
$g_x$ & $g_{[-, q, 1]}$ \\
$e_y$ & $e_{[-, q, 2]}$ \\
$f_x$ & $f_{[-, q, 1]}$ \\
$\alpha_{z}$ & $Chord(C,C_2,E)_{[-, q, 1]}$ \\
$\beta_{z}$ & $Chord(D,F,A)_{[-, q, 1]}$ \\
$\gamma_{z}$ & $Chord(A,C,F)_{[-, q, 1]}$ \\
$\delta_{z}$ & $Chord(A,C,E)_{[-, q, 1]}$ \\
$\epsilon_{w}$ & $Chord(A,C,F)_{[-, h, 1]}$ \\
$\zeta_{w}$ & $Chord(F,A,C)_{[-, h, 1]}$ \\
\end{tabular}
\caption{Mapping of terminal symbols to musical feature vectors of $G_1$.}
\label{Tb1}
\end{minipage}
\hfill
\begin{minipage}{0.48\textwidth}
\centering
\begin{tabular}{c|c}
Symbol & Note \\
\hline
$c_{v}$ & $c_{[-, e, 1]}$ \\
$g_{v}$ & $g_{[-, e, 1]}$ \\
$r_{t}$ & $r_{[-, e, 1]}$ \\
$a_{u}$ & $a_{[\flat, e, 1]}$ \\
$e_{v}$ & $e_{[\flat, e, 1]}$ \\
$r_{t}$ & $r_{[-, e, 1]}$ \\
\end{tabular}
\caption{Mapping of terminal symbols to musical feature vectors of $G_2$.}
\label{Tb2}
\end{minipage}
\end{table}

For this $G_s$, we can create the following derivation steps:
\begin{itemize}
    \item{$(S_1, S_2) \Rightarrow^1 (AABA, AABA) \Rightarrow^2 (MHMHBMH, PLPLBPL) \\ 
    \Rightarrow^3 (MHMHM_1H_1MH, PLPLPLPL)
    \\ \Rightarrow^4 (c_{y}c_{y}a_{x}g_{x}Hc_{y}c_{y}a_{x}g_{x}HM_1H_1c_{y}c_{y}a_{x}g_{x}H, 
    \\ c_{v}g_{v}c_{v}g_{v}r_{v}g_{v}c_{v}g_{v}Lc_{v}g_{v}c_{v}g_{v}r_{v}g_{v}c_{v}g_{v}LPLc_{v}g_{v}c_{v}g_{v}r_{v}g_{v}c_{v}g_{v}L)
    \\ \Rightarrow \hdots$}
\end{itemize}

\begin{figure}[H]
\centering
\includegraphics[scale=0.32]{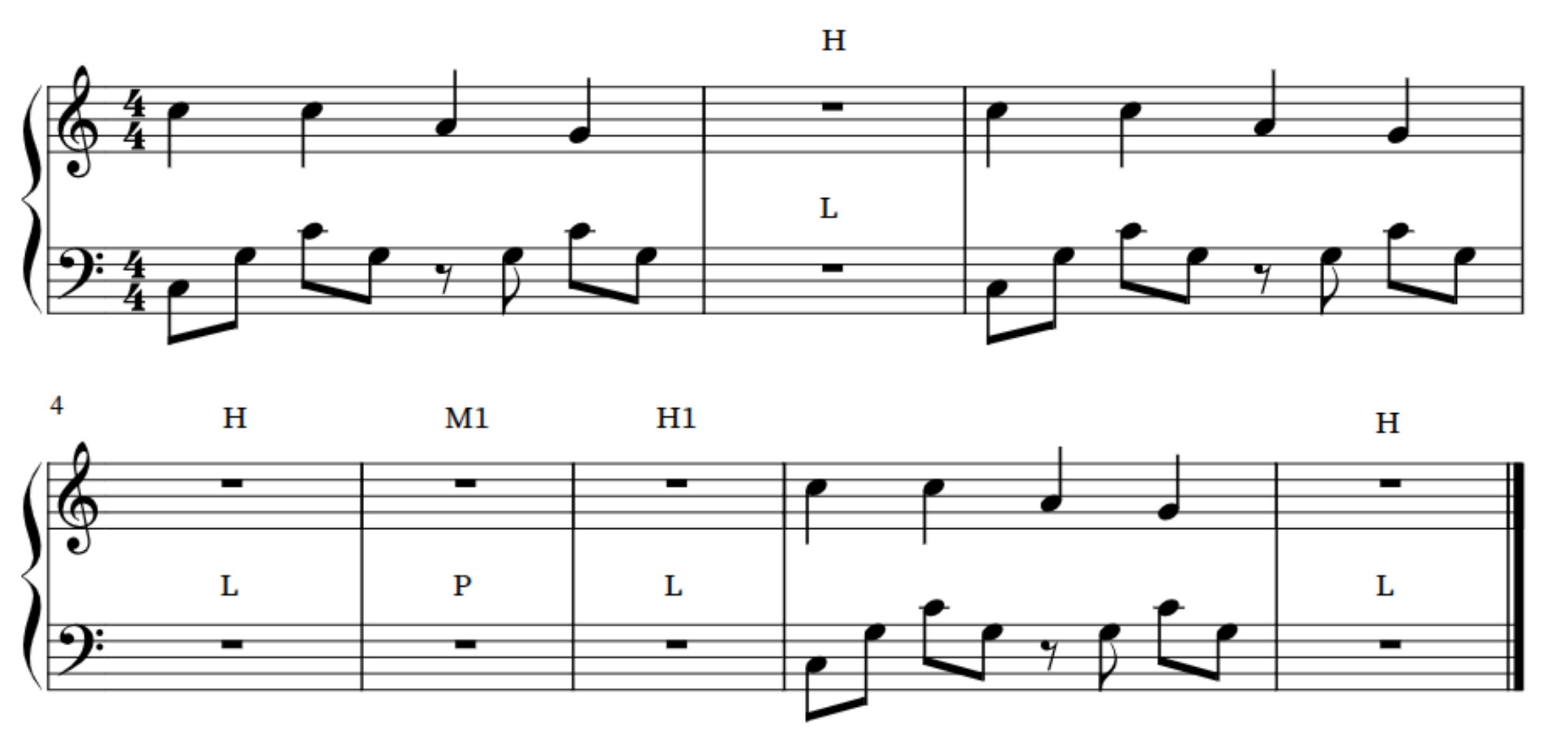}
\[
\Rightarrow^5
\]
\includegraphics[scale=0.32]{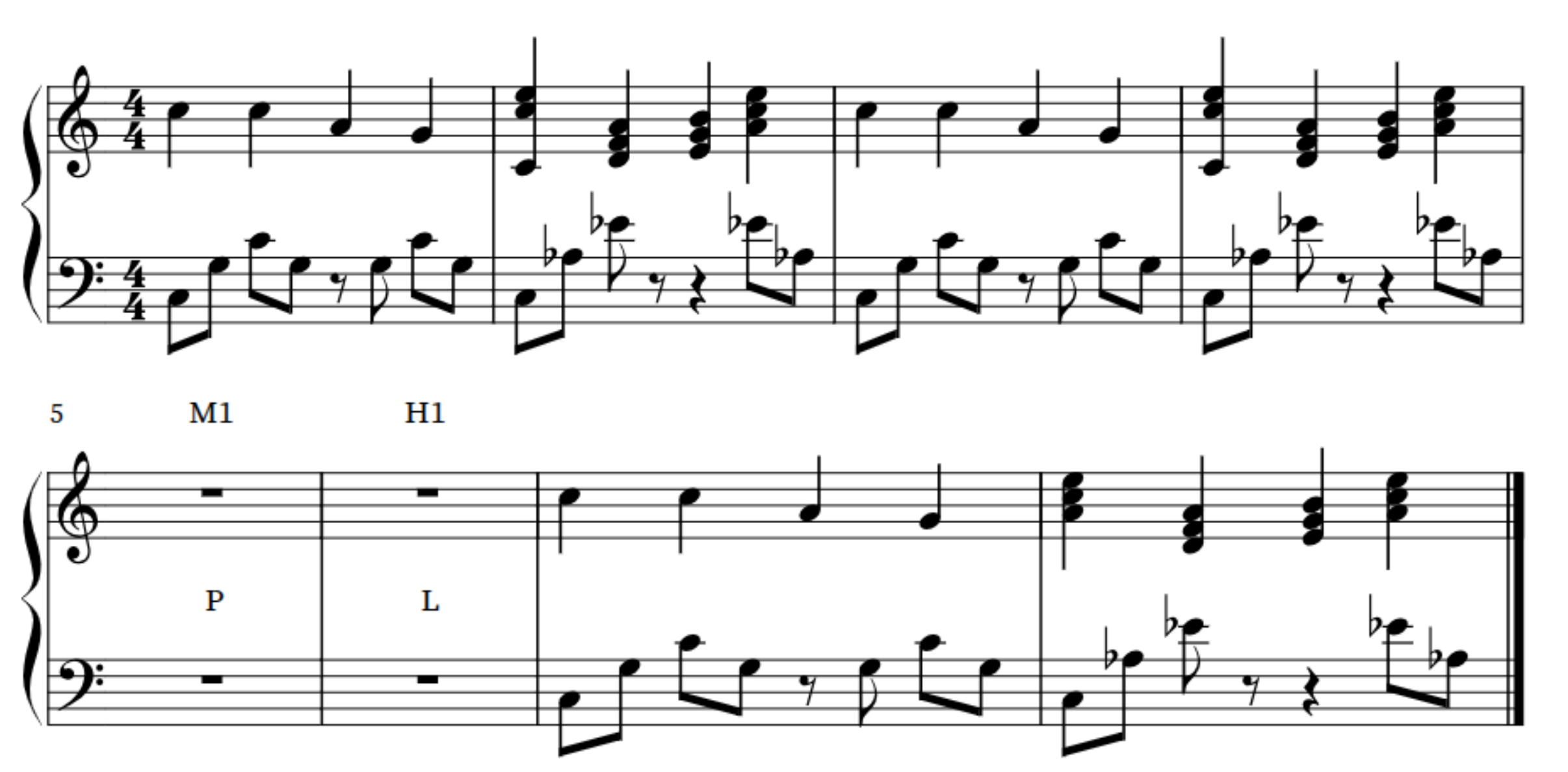}
\[
\Rightarrow^6
\]
\includegraphics[scale=0.32]{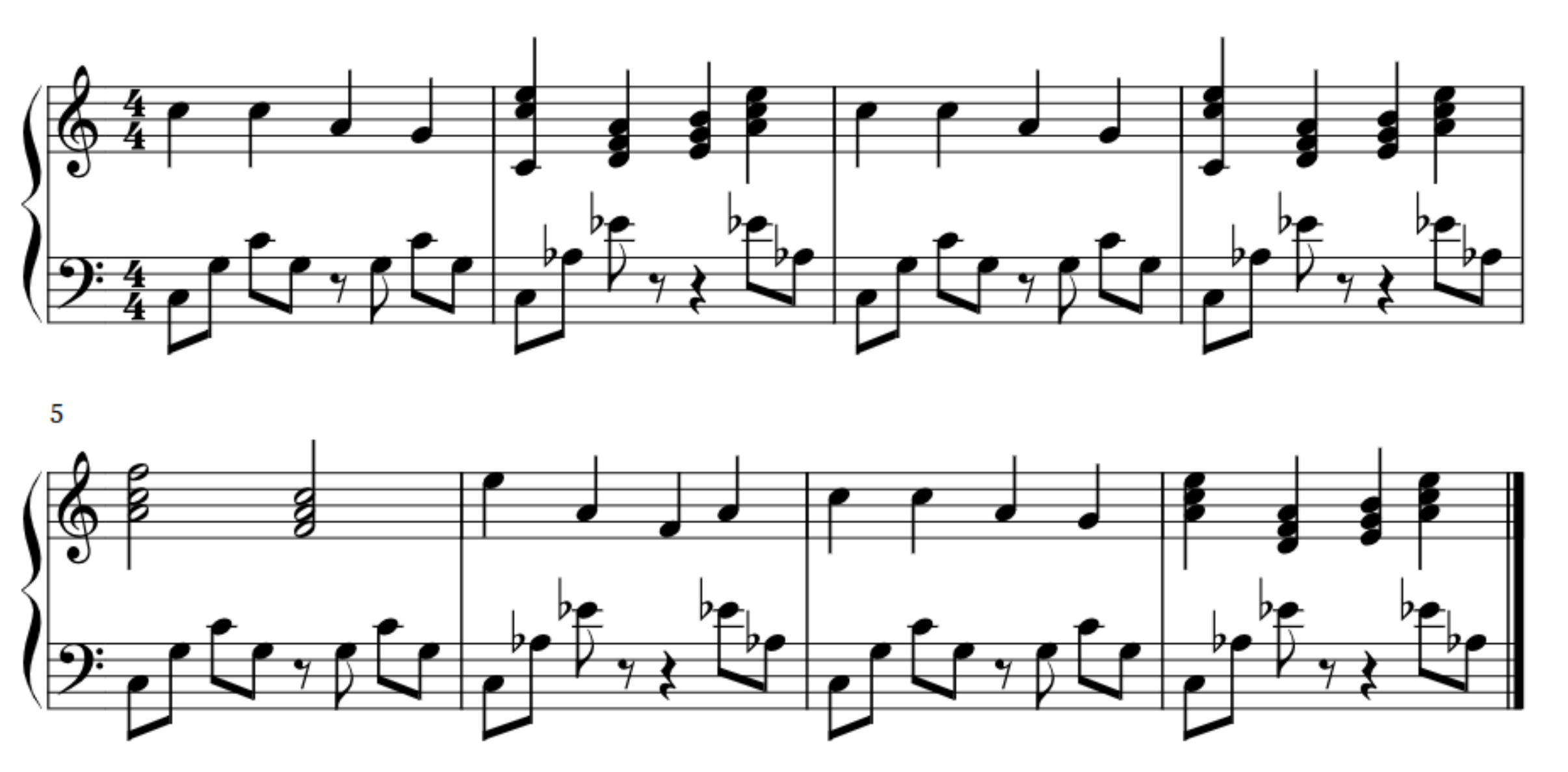}
\caption{The fifth and sixth derivation step shown in music staff that corresponds to $(5,5) \in Q$, and $(7,7) \in Q$.}
\label{fig:def1final1}
\end{figure}

Instead of writing out terminal symbols, it is much more interesting to demonstrate terminal symbols already in the music staff. Nonterminal symbols are blank bars that represent the structure. Fig.~\ref{fig:def1final1} describes the correspondence between the fifth and sixth derivation steps in $G_s$ and their musical interpretation. More specifically, during $\Rightarrow^5$, $G_s$ rewrites nonterminals $H$ and $L$ from $G_1$ and $G_2$, respectively; as a result, all A parts are completed.  During $\Rightarrow^6$, $G_s$ completes the generation of the sentence and, therefore, its corresponding musical piece by filling in the missing part of the generated score.

\section{Example}
\label{sec:Example}

Until now, we have been generating music for only one instrument. Finally, we will show how our model could generate jazz music. This music is going to be interpreted by a piano and saxophone. The music will take jazz from AABA and will be generated in three strings, two for piano and one for saxophone. So far, we have used variation, tone duration, and tone octave for our generated tokens. Now, we will also incorporate dynamics. An example of a grammar system generating such computation follows: 

$$G_s = (G_1, G_2, G_3, Q),$$ in which 
\begin{itemize}
    \item{$G_1 = (\{S_1, M_1, M_2, A, B, N\},\, \{f_{[-, q, 1, -]}, c_{[-, q, 1, -]}, f_{[\downarrow, q, 1, p]}, c_{[\downarrow, q, 1, p]}, g_{[-, q, 2, -]},
    \\ d_{[-, q, 2, -]}, g_{[\downarrow, q, 2, p]}, d_{[\downarrow, q, 2, p]}, e_{[-, h, 2, -]}, g_{[-, h, 2, -]}, e_{[\downarrow, h, 2, p]}, g_{[\downarrow, h, 2, p]}, f_{[-, h, 2, -]}, 
    \\ a_{[-, h, 2, -]}, f_{[\downarrow, h, 2, p]}, a_{[\downarrow, h, 2, p]}\},\, 
    \\ \{1\colon S \rightarrow (AABA),
    \\ 2\colon (A, A, A) \rightarrow (M_1M_2M_2M_1, M_1M_2M_2M_1, M_1M_2M_2M_1),
    \\ 3\colon (M_1, M_1, M_1) \rightarrow (f_{[-, q, 1, -]}c_{[-, q, 1, -]}c_{[-, h, 1, -]}, 
    \\f_{[\downarrow, q, 1, p]}c_{[\downarrow, q, 1, p]}c_{[\downarrow, h, 1, p]}, f_{[-, q, 1, -]}c_{[-, q, 1, -]}c_{[-, h, 2, -]}),
    \\ 4\colon (M_1, M_1, M_1) \rightarrow (g_{[-, q, 2, -]}d_{[-, q, 2, -]}d_{[-, h, 2, -]}, 
    \\g_{[\downarrow, q, 2, p]}d_{[\downarrow, q, 2, p]}d_{[\downarrow, h, 2, p]}, g_{[-, q, 2, -]}d_{[-, q, 2, -]}d_{[-, h, 2, -]}),
    \\ 5\colon (M_2, M_2, M_2) \rightarrow (e_{[-, h, 2, -]}g_{[-, h, 2, -]}, e_{[\downarrow, h, 2, p]}g_{[\downarrow, h, 2, p]}, e_{[-, h, 2, -]}g_{[-, h, 2, -]}),
    \\ 6\colon (M_2, M_2, M_2) \rightarrow (f_{[-, h, 2, -]}a_{[-, h, 2, -]}, f_{[\downarrow, h, 2, p]}a_{[\downarrow, h, 2, p]}, f_{[-, h, 2, -]}a_{[-, h, 2, -]}),
    \\ 7\colon B \rightarrow (NNNN)
    \\ 8\colon N \rightarrow (r_{[-, f, -, -]})
    \})$,}
    \item{$G_2 = (\{S_2, A, B, P, R, N\},\, \{\gamma_{[-, h, 1, -]}, \gamma_{[P, h, 1, -]}, \gamma_{[R, h, 1, p]}, r_{[-, f, -, -]}\},\, 
    \\ \{1\colon S \rightarrow (AABA),
    \\ 2\colon (A, A, A) \rightarrow (PRPR, PRPR, PRPR), 
    \\ 3\colon (P, P, P) \rightarrow (\gamma_{[-, h, 1, -]}\gamma_{[P, h, 1, -]}R, \gamma_{[-, h, 1, p]}\gamma_{[P, h, 1, p]}R, \gamma_{[-, h, 1, -]}\gamma_{[P, h, 1, -]}R),
    \\ 4\colon (R, R, R) \rightarrow (\gamma_{[-, h, 1, -]}\gamma_{[R, h, 1, -]}P, \gamma_{[-, h, 1, p]}\gamma_{[R, h, 1, p]}P, \gamma_{[-, h, 1, -]}\gamma_{[R, h, 1, -]}P),
    \\ 5\colon (P, P, P) \rightarrow (\gamma_{[-, h, 1, -]}\gamma_{[P, h, 1, -]}, \gamma_{[-, h, 1, p]}\gamma_{[P, h, 1, p]}, \gamma_{[-, h, 1, -]}\gamma_{[P, h, 1, -]}),
    \\ 6\colon (R, R, R) \rightarrow (\gamma_{[-, h, 1, -]}\gamma_{[R, h, 1, -]}, \gamma_{[-, h, 1, p]}\gamma_{[R, h, 1, p]}, \gamma_{[-, h, 1, -]}\gamma_{[R, h, 1, -]}),
    \\ 7\colon B \rightarrow (NNNN)
    \\ 8\colon N \rightarrow (r_{[-, f, -, -]})
    \})$,}
    \item{$G_3 = (\{S_3, A, B, H, M_{31}, M_{32}\},\, \{e_{[-, h, 1, -]}, g_{[-, h, 1, -]}, a_{[-, h, 1, -]}, f_{[-, h, 1, -]}, \\ h_{[-, h, 1, -]}, c_{[-, q, 1, -]}\},\, 
    \\ \{1\colon S \rightarrow (AABA),
    \\ 2\colon (A, A, A) \rightarrow (MMMM, MMMM, MMMM),
    \\ 3\colon (M, M, M) \rightarrow (e_{[-, h, 1, -]}g_{[-, h, 1, -]}, e_{[\uparrow, h, 1, -]}g_{[\uparrow, h, 1, -]}, e_{[-, h, 1, -]}g_{[-, h, 1, -]}),
    \\ 4\colon (M, M, M) \rightarrow (a_{[-, h, 1, -]}f_{[-, h, 1, -]}, a_{[\uparrow, h, 1, -]}f_{[\uparrow, h, 1, -]}, a_{[-, h, 1, -]}f_{[-, h, 1, -]}),
    \\ 5\colon (M, M, M) \rightarrow (g_{[-, h, 1, -]}e_{[-, h, 1, -]}, g_{[\uparrow, h, 1, -]}e_{[\uparrow, h, 1, -]}, g_{[-, h, 1, -]}e_{[-, h, 1, -]}),
    \\ 6\colon (M, M, M) \rightarrow (f_{[-, h, 1, -]}f_{[-, h, 1, -]}, f_{[\uparrow, h, 1, -]}f_{[\uparrow, h, 1, -]}, f_{[-, h, 1, -]}f_{[-, h, 1, -]}),
    \\ 7\colon B \rightarrow (HM_{31}M_{32}H)
    \\ 8\colon (H, H) \rightarrow (\alpha_{[-, h, 1, -]}\beta_{[-, h, 1, -]}, \alpha_{[r, h, 1, -]}\beta_{[r, h, 1, -]})
    \\ 9\colon M_{31} \rightarrow (e_{[-, h, 1, -]}g_{[-, h, 1, -]}a_{[-, h, 1, -]}h_{[-, h, 1, -]})
    \\ 10\colon M_{32} \rightarrow (h_{[-, q, 1, -]}c_{[-, q, 1, -]}a_{[-, q, 1, -]}f_{[-, q, 1, -]})
    \})$,}
    \item{$Q = \{(1, 1, 1), (2, 2, 2), (3, 3, 3), (3, 5, 3), (4, 4, 4), (4, 6, 4), (5, 4, 5), (5, 6, 5), \\ (6, 3, 6), (6, 5, 6), (7, 7, 7), (8, 8, 8), (8, 8, 9), (8, 8, 10)\}$.}
\end{itemize}

\begin{figure}[H]
\centering
\includegraphics[scale=0.45]{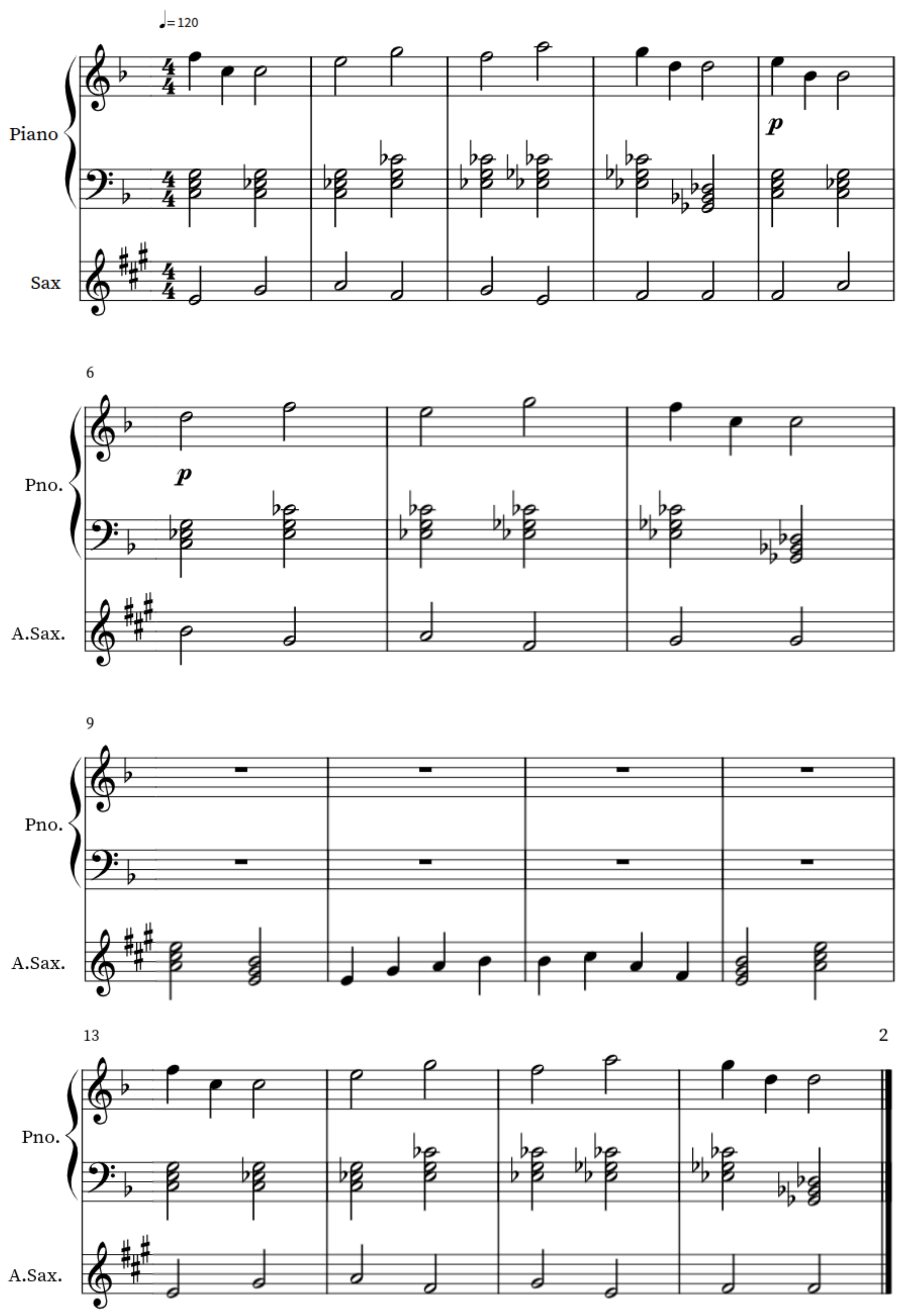}
\caption{Illustrative example of multi-instrument jazz composition.}
\label{fig:JazzExamples}
\end{figure}

A composition that could be generated by the presented grammar system is shown in Fig. \ref{fig:JazzExamples}. It shows that grammar can generate meaningful music with various music techniques. To describe what is in the figure, we would start with the piano part. In the piano part, the A section of the composition presents the main theme and completes the harmony in the treble clef, while additional harmonic support is found in the bass clef. Alongside the piano, the saxophone is there to provide a second harmonic party to enrich the melody. The role of the Sax is to create an interesting contrast to the main melody. While the primary theme ascends, the Sax line moves downwards, which creates a playful tension and enriches the overall texture. A bridge is created by Sax solo, which is an alternation between harmonic and melodic material to create contrast with the A sections and a bridge between the piano part of the main theme and the last repetition of the main theme that ends the composition. 

Tables \ref{Tb3} and \ref{Tb4} show interpretation of symbols from $G_s$ in this section. 
\begin{table}[H]
\centering
\begin{minipage}{0.48\textwidth}
\centering
\begin{tabular}{c|c}
Symbol & Note \\
\hline
$\gamma$ & $Chord(C,E,G)$ \\
$\gamma$ & $Chord(C,Es,G)$ \\
$\gamma$ & $Chord(Es,G,Ces)$ \\
$\gamma$ & $Chord(Es,Ges,Ces)$ \\
$\gamma$ & $Chord(Ges,Hes,Des)$ \\
\end{tabular}
\caption{Mapping of terminal symbols to chords from Tonnetz \cite{gogins2006score} walk using PR transformations of $G_2$.}
\label{Tb3}
\end{minipage}
\hfill
\begin{minipage}{0.48\textwidth}
\centering
\begin{tabular}{c|c}
Symbol & Chord \\
\hline
$\alpha$ & $Chord(A,C,E)$ \\
$\beta$ & $Chord(E,G,H)$ \\
\end{tabular}
\caption{Mapping of terminal symbols to musical feature vectors of $G_3$.}
\label{Tb4}
\end{minipage}
\end{table}

To see more song examples and implementation details visit our GitHub repository.\footnote{\label{github}Implementation details at \href{https://github.com/NaKamize/music-grammar-system}{https://github.com/NaKamize/music-grammar-system}}

\section{Evaluation}
\label{sec:Evaluation}
We mentioned that music is a creative process, and because of that, it is difficult to find a mathematical formula that provides a number or graph to help compare our method to existing algorithms for music generation. And we don't need that. The biggest advantage is the enforcement of the rules and their synchronization, which allows the music structure to fit its nature perfectly. We showed this through the provided examples. Generated examples keep the musical structure as it was intended and follow the rules of music theory. This is due to the correctly selected rules. The playable sound examples are stored in GitHub\footnoteref{github} with the implementation and implementation details. 

To compare our method to L-systems, we are able to generate not just the fractal music but any music that has structure. We don't require postprocessing of the generated string; it can be interpreted instantly. Probabilistic formal models have the advantage that they can learn to imitate any style and generate that style of music. In comparison, our method is as good as the person who is creating the rules. The tone rules have to fit a specific style or melody.

This method is great at creating synchronized multi-instrument pieces, and its use could be in the procedural generation of music for computer games, as \cite{eibensteiner2018procedural}. There have been several attempts to enhance music generation using neural networks. However, they often struggle to capture long-term dependencies or musical structure. A hybrid approach that combines them with our model could be advantageous. Those approaches keep the rich and expressive sound of neural networks and combine it with the needed structure and dependencies.

\section{Conclusion}
\label{sec:Conclusion}

To  summarize the present application-oriented paper as simply as possible, we have demonstrated how to orchestrate music by using grammar systems (see Section~\ref{sec:Definitions} and \ref{sec:Orchestration}). In addition, we have illustrated an orchestration of this kind by an example (see Section \ref{sec:Example}). 

Although we have described this kind of orchestration in a rather great detail, there still remain many open problem areas related to the subject of this paper. Next, we suggest five of them.

(1) Investigate classical topics of formal language theory, such as decidable problems or closure properties, in terms of the systems from Section~\ref{sec:Definitions}. 

(2) Conceptualize, re-formulate  and investigate the subject of this paper in terms of other language models, such as jumping or regulated grammars and automata (see~\cite{meduna2024jumping, Me14}).   

(3) Restrict the systems from Section~\ref{sec:Definitions} so they can use only context-free or even linear rules.  What kind of music can be orchestrated by systems restricted in this way?

(4) Many compositions for orchestras frequently contain long musical passages during which several instruments simultaneously play the same music. Can the grammar systems considered in Section~\ref{sec:Definitions} be modified so that a single component produce a score for all these instruments, which play the same music? Even more generally, can these systems be modified so that a single component produces scores for several instruments, possibly playing different music? 

(5) Consider only smaller-sized orchestras, such as chamber orchestras. What are the simplest possible versions of the grammar systems that can orchestrate them? 

\section*{Acknowledgments}
This work was supported by Brno University of Technology grant FIT-S-23-8209.

%

\bibliographystyle{eptcs}
\bibliography{paper}

\begin{thebibliography}{10}
\providecommand{\bibitemdeclare}[2]{}
\providecommand{\surnamestart}{}
\providecommand{\surnameend}{}
\providecommand{\urlprefix}{Available at }
\providecommand{\url}[1]{\texttt{#1}}
\providecommand{\href}[2]{\texttt{#2}}
\providecommand{\urlalt}[2]{\href{#1}{#2}}
\providecommand{\doi}[1]{doi:\urlalt{https://doi.org/#1}{#1}}
\providecommand{\eprint}[1]{arXiv:\urlalt{https://arxiv.org/abs/#1}{#1}}
\providecommand{\bibinfo}[2]{#2}

\bibitemdeclare{book}{adamatzky2016designing}
\bibitem{adamatzky2016designing}
\bibinfo{editor}{Andrew \surnamestart Adamatzky\surnameend} \&
  \bibinfo{editor}{Genaro~J. \surnamestart Martínez\surnameend}, editors
  (\bibinfo{year}{2016}): \emph{\bibinfo{title}{Designing Beauty: The Art of
  Cellular Automata}}, \bibinfo{edition}{1st} edition.
\newblock {\slshape \bibinfo{series}{Emergence, Complexity and
  Computation}}~\bibinfo{volume}{20}, \bibinfo{publisher}{Springer},
  \doi{10.1007/978-3-319-32922-7}.
\newblock \bibinfo{note}{Kindle Edition}.

\bibitemdeclare{book}{AU}
\bibitem{AU}
\bibinfo{author}{A.V. \surnamestart Aho\surnameend} \& \bibinfo{author}{J.D.
  \surnamestart Ullman\surnameend} (\bibinfo{year}{1972}):
  \emph{\bibinfo{title}{The Theory of Parsing, Translation, and Compiling}}.
\newblock \bibinfo{publisher}{Prentice-Hall, Series in Automatic Computation}.

\bibitemdeclare{article}{albarracin2021using}
\bibitem{albarracin2021using}
\bibinfo{author}{David~D. \surnamestart Albarracín-Molina\surnameend},
  \bibinfo{author}{Alfredo \surnamestart Raglio\surnameend},
  \bibinfo{author}{Francisco \surnamestart Rivas-Ruiz\surnameend} \&
  \bibinfo{author}{Francisco~J. \surnamestart Vico\surnameend}
  (\bibinfo{year}{2021}): \emph{\bibinfo{title}{Using Formal Grammars as
  Musical Genome}}.
\newblock {\slshape \bibinfo{journal}{Applied Sciences}}
  \bibinfo{volume}{11}(\bibinfo{number}{9}), p. \bibinfo{pages}{4151},
  \doi{10.3390/app11094151}.

\bibitemdeclare{incollection}{bel1992modelling}
\bibitem{bel1992modelling}
\bibinfo{author}{Bernard \surnamestart Bel\surnameend} \& \bibinfo{author}{Jim
  \surnamestart Kippen\surnameend} (\bibinfo{year}{1992}):
  \emph{\bibinfo{title}{Modelling music with grammars: formal language
  representation in the {Bol} Processor}}.
\newblock In: {\slshape \bibinfo{booktitle}{Computer Representations and Models
  in Music}}, \bibinfo{publisher}{Academic Press}, pp.
  \bibinfo{pages}{207--238}.
\newblock \urlprefix\url{https://shs.hal.science/halshs-00004506}.

\bibitemdeclare{article}{edwards2011algorithmic}
\bibitem{edwards2011algorithmic}
\bibinfo{author}{Michael \surnamestart Edwards\surnameend}
  (\bibinfo{year}{2011}): \emph{\bibinfo{title}{Algorithmic Composition:
  Computational Thinking in Music}}.
\newblock {\slshape \bibinfo{journal}{Communications of the ACM}}
  \bibinfo{volume}{54}(\bibinfo{number}{7}), pp. \bibinfo{pages}{58--67},
  \doi{10.1145/1965724.1965742}.

\bibitemdeclare{inproceedings}{eibensteiner2018procedural}
\bibitem{eibensteiner2018procedural}
\bibinfo{author}{Lukas \surnamestart Eibensteiner\surnameend}
  (\bibinfo{year}{2018}): \emph{\bibinfo{title}{Procedural Music Generation
  with Grammars}}.
\newblock In: {\slshape \bibinfo{booktitle}{Proceedings of the 22nd Central
  European Seminar on Computer Graphics (CESCG)}}.

\bibitemdeclare{inproceedings}{gilbert2007probabilistic}
\bibitem{gilbert2007probabilistic}
\bibinfo{author}{\surnamestart Édouard Gilbert\surnameend} \&
  \bibinfo{author}{Darrell \surnamestart Conklin\surnameend}
  (\bibinfo{year}{2007}): \emph{\bibinfo{title}{A Probabilistic Context-Free
  Grammar for Melodic Reduction}}.
\newblock In: {\slshape \bibinfo{booktitle}{Proceedings of the International
  Workshop on Artificial Intelligence and Music}}, \bibinfo{address}{Hyderabad,
  India}, pp. \bibinfo{pages}{83--94}.

\bibitemdeclare{inproceedings}{gogins2006score}
\bibitem{gogins2006score}
\bibinfo{author}{Michael \surnamestart Gogins\surnameend}
  (\bibinfo{year}{2006}): \emph{\bibinfo{title}{Score Generation in
  Voice-Leading and Chord Spaces}}.
\newblock In \bibinfo{editor}{Georg \surnamestart Essl\surnameend} \&
  \bibinfo{editor}{Ichiro \surnamestart Fujinaga\surnameend}, editors:
  {\slshape \bibinfo{booktitle}{Proceedings of the 2006 International Computer
  Music Conference (ICMC)}}, \bibinfo{publisher}{International Computer Music
  Association}, pp. \bibinfo{pages}{455--457}.

\bibitemdeclare{book}{MAH}
\bibitem{MAH}
\bibinfo{author}{M.~A. \surnamestart Harrison\surnameend}
  (\bibinfo{year}{1978}): \emph{\bibinfo{title}{Introduction to Formal Language
  Theory}}.
\newblock \bibinfo{publisher}{Addison-Wesley Longman Publishing Co., Inc.},
  \bibinfo{address}{Boston, MA, USA}.

\bibitemdeclare{article}{humphreys2021investigation}
\bibitem{humphreys2021investigation}
\bibinfo{author}{David \surnamestart Humphreys\surnameend},
  \bibinfo{author}{Kirill \surnamestart Sidorov\surnameend},
  \bibinfo{author}{Andrew \surnamestart Jones\surnameend} \&
  \bibinfo{author}{David \surnamestart Marshall\surnameend}
  (\bibinfo{year}{2021}): \emph{\bibinfo{title}{An Investigation of Music
  Analysis by the Application of Grammar-Based Compressors}}.
\newblock {\slshape \bibinfo{journal}{Journal of New Music Research}}
  \bibinfo{volume}{50}(\bibinfo{number}{4}), pp. \bibinfo{pages}{312--341},
  \doi{10.1080/09298215.2021.1978505}.

\bibitemdeclare{inproceedings}{jin2013formal}
\bibitem{jin2013formal}
\bibinfo{author}{Zeyu \surnamestart Jin\surnameend} \&
  \bibinfo{author}{Roger~B. \surnamestart Dannenberg\surnameend}
  (\bibinfo{year}{2013}): \emph{\bibinfo{title}{Formal Semantics for Music
  Notation Control Flow}}.
\newblock In: {\slshape \bibinfo{booktitle}{Proceedings of the International
  Computer Music Conference (ICMC)}}.
\newblock \urlprefix\url{http://hdl.handle.net/2027/spo.bbp2372.2013.010}.

\bibitemdeclare{inproceedings}{jurish2004music}
\bibitem{jurish2004music}
\bibinfo{author}{Bryan \surnamestart Jurish\surnameend} (\bibinfo{year}{2004}):
  \emph{\bibinfo{title}{Music as a Formal Language}}.
\newblock In \bibinfo{editor}{Fränk \surnamestart Zimmer\surnameend}, editor:
  {\slshape \bibinfo{booktitle}{bang | pure data}}, \bibinfo{publisher}{Wolke
  Verlag}, \bibinfo{address}{Hofheim}, pp. \bibinfo{pages}{45--58}.

\bibitemdeclare{misc}{Kadlec2022}
\bibitem{Kadlec2022}
\bibinfo{author}{Tim \surnamestart Kadlec\surnameend}, \bibinfo{author}{Ivica
  \surnamestart Gabrišová\surnameend}, \bibinfo{author}{Janka \surnamestart
  Jámborová\surnameend}, \bibinfo{author}{Michal \surnamestart
  Vojáček\surnameend}, \bibinfo{author}{Emily \surnamestart
  Beynon\surnameend}, \bibinfo{author}{Robert \surnamestart Heger\surnameend}
  \& \bibinfo{author}{Halka \surnamestart Klánská\surnameend}
  (\bibinfo{year}{2022}): \emph{\bibinfo{title}{Methodology: Increasing the
  Efficiency and Quality of Instrumentalists' Preparation for Orchestral
  Auditions}}.
\newblock \bibinfo{note}{Accessed: 2025-03-14}.

\bibitemdeclare{inproceedings}{keller2007grammatical}
\bibitem{keller2007grammatical}
\bibinfo{author}{Robert~M. \surnamestart Keller\surnameend} \&
  \bibinfo{author}{David~R. \surnamestart Morrison\surnameend}
  (\bibinfo{year}{2007}): \emph{\bibinfo{title}{A Grammatical Approach to
  Automatic Improvisation}}.
\newblock In: {\slshape \bibinfo{booktitle}{Proceedings of the 4th Sound and
  Music Computing Conference (SMC)}}, \bibinfo{address}{Lefkada, Greece}, pp.
  \bibinfo{pages}{330--337}.

\bibitemdeclare{phdthesis}{lukas2006multigenerative}
\bibitem{lukas2006multigenerative}
\bibinfo{author}{Roman \surnamestart Lukáš\surnameend}
  (\bibinfo{year}{2006}): \emph{\bibinfo{title}{Multigenerative Grammar
  Systems}}.
\newblock \bibinfo{type}{Ph.d. dissertation}, \bibinfo{school}{Brno University
  of Technology}, \bibinfo{address}{Brno, Czech Republic}.
\newblock \bibinfo{note}{Supervisor: Prof. RNDr. Alexander Meduna, CSc.}

\bibitemdeclare{mastersthesis}{manousakis2006musical}
\bibitem{manousakis2006musical}
\bibinfo{author}{Stelios \surnamestart Manousakis\surnameend}
  (\bibinfo{year}{2006}): \emph{\bibinfo{title}{Musical L-Systems}}.
\newblock \bibinfo{type}{Master's thesis}, \bibinfo{school}{Royal Conservatory,
  The Hague}.

\bibitemdeclare{article}{manousakis2009non}
\bibitem{manousakis2009non}
\bibinfo{author}{Stelios \surnamestart Manousakis\surnameend}
  (\bibinfo{year}{2009}): \emph{\bibinfo{title}{Non-Standard Sound Synthesis
  with L-Systems}}.
\newblock {\slshape \bibinfo{journal}{Leonardo Music Journal}}
  \bibinfo{volume}{19}, pp. \bibinfo{pages}{85--94},
  \doi{10.1162/lmj.2009.19.85}.

\bibitemdeclare{incollection}{mccormack1996grammar}
\bibitem{mccormack1996grammar}
\bibinfo{author}{Jon \surnamestart McCormack\surnameend}
  (\bibinfo{year}{1996}): \emph{\bibinfo{title}{Grammar-Based Music
  Composition}}.
\newblock In \bibinfo{editor}{A.~\surnamestart Stocker\surnameend},
  \bibinfo{editor}{M.~\surnamestart Schenker\surnameend} \&
  \bibinfo{editor}{L.~M. \surnamestart Browne\surnameend}, editors: {\slshape
  \bibinfo{booktitle}{Complex Systems: From Local Interactions to Global
  Phenomena}}, \bibinfo{volume}{96}, \bibinfo{publisher}{IOS Press}, pp.
  \bibinfo{pages}{321--336}.

\bibitemdeclare{book}{meduna2020handbook}
\bibitem{meduna2020handbook}
\bibinfo{author}{Alexander \surnamestart Meduna\surnameend},
  \bibinfo{author}{Petr \surnamestart Horáček\surnameend} \&
  \bibinfo{author}{Martin \surnamestart Tomko\surnameend}
  (\bibinfo{year}{2020}): \emph{\bibinfo{title}{Handbook of Mathematical Models
  for Languages and Computation}}.
\newblock \bibinfo{series}{Computing and Networks}, \bibinfo{publisher}{The
  Institution of Engineering and Technology}, \bibinfo{address}{London, UK}.
\newblock \bibinfo{note}{Kindle Edition}.

\bibitemdeclare{book}{meduna2024jumping}
\bibitem{meduna2024jumping}
\bibinfo{author}{Alexander \surnamestart Meduna\surnameend} \&
  \bibinfo{author}{Zbyněk \surnamestart Křivka\surnameend}
  (\bibinfo{year}{2024}): \emph{\bibinfo{title}{Jumping Computation: Updating
  Automata and Grammars for Discontinuous Information Processing}}.
\newblock \bibinfo{publisher}{CRC Press}, \bibinfo{address}{Boca Raton}.

\bibitemdeclare{book}{Me14}
\bibitem{Me14}
\bibinfo{author}{Alexander \surnamestart Meduna\surnameend} \&
  \bibinfo{author}{Petr \surnamestart Zemek\surnameend} (\bibinfo{year}{2014}):
  \emph{\bibinfo{title}{Regulated Grammars and Automata}}.
\newblock \bibinfo{publisher}{Springer}, \bibinfo{address}{New York},
  \doi{10.1007/978-1-4939-0369-6}.

\bibitemdeclare{inproceedings}{melkonian2019music}
\bibitem{melkonian2019music}
\bibinfo{author}{Orestis \surnamestart Melkonian\surnameend}
  (\bibinfo{year}{2019}): \emph{\bibinfo{title}{Music as Language: Putting
  Probabilistic Temporal Graph Grammars to Good Use}}.
\newblock In: {\slshape \bibinfo{booktitle}{Proceedings of the 7th ACM SIGPLAN
  International Workshop on Functional Art, Music, Modeling, and Design (FARM
  2019)}}, \bibinfo{publisher}{Association for Computing Machinery}, pp.
  \bibinfo{pages}{1--10}, \doi{10.1145/3331543.3342576}.

\bibitemdeclare{misc}{PankhurstSonataForm}
\bibitem{PankhurstSonataForm}
\bibinfo{author}{Tom \surnamestart Pankhurst\surnameend}:
  \emph{\bibinfo{title}{Sonata Form}}.
\newblock
  \bibinfo{howpublished}{\url{https://alevelmusic.com/alevelcompositionhelp
  /composing-help/sonata-form-2/sonata-form/}}.
\newblock \bibinfo{note}{Accessed: 2025-03-15}.

\bibitemdeclare{inproceedings}{prusinkiewicz1986score}
\bibitem{prusinkiewicz1986score}
\bibinfo{author}{Przemyslaw \surnamestart Prusinkiewicz\surnameend}
  (\bibinfo{year}{1986}): \emph{\bibinfo{title}{Score generation with
  L-systems}}.
\newblock In: {\slshape \bibinfo{booktitle}{Proceedings of the International
  Computer Music Conference (ICMC)}}, pp. \bibinfo{pages}{455--457}.

\bibitemdeclare{misc}{ravelliJazzForm2025}
\bibitem{ravelliJazzForm2025}
\bibinfo{author}{Leonardo \surnamestart Ravelli\surnameend}
  (\bibinfo{year}{2025}): \emph{\bibinfo{title}{Understanding music to
  improvise better: Form in jazz standards}}.
\newblock \bibinfo{note}{Accessed: 2025-03-14}.

\bibitemdeclare{phdthesis}{krakowski2009rhythmically}
\bibitem{krakowski2009rhythmically}
\bibinfo{author}{Sergio Krakowski~Costa \surnamestart Rego\surnameend}
  (\bibinfo{year}{2009}): \emph{\bibinfo{title}{Rhythmically-Controlled
  Automata Applied to Musical Improvisation}}.
\newblock \bibinfo{type}{Ph.d. dissertation}, \bibinfo{school}{Instituto
  Nacional de Matemática Pura e Aplicada (IMPA)}, \bibinfo{address}{Rio de
  Janeiro, Brazil}.

\bibitemdeclare{inproceedings}{rodrigues2016evolving}
\bibitem{rodrigues2016evolving}
\bibinfo{author}{Ana \surnamestart Rodrigues\surnameend},
  \bibinfo{author}{Ernesto \surnamestart Costa\surnameend},
  \bibinfo{author}{Amílcar \surnamestart Cardoso\surnameend},
  \bibinfo{author}{Penousal \surnamestart Machado\surnameend} \&
  \bibinfo{author}{Tiago \surnamestart Cruz\surnameend} (\bibinfo{year}{2016}):
  \emph{\bibinfo{title}{Evolving L-Systems with Musical Notes}}.
\newblock In \bibinfo{editor}{Colin \surnamestart Johnson\surnameend},
  \bibinfo{editor}{Alvaro \surnamestart Carballal\surnameend} \&
  \bibinfo{editor}{João \surnamestart Correia\surnameend}, editors: {\slshape
  \bibinfo{booktitle}{Evolutionary and Biologically Inspired Music, Sound, Art
  and Design}}, {\slshape \bibinfo{series}{Lecture Notes in Computer Science}}
  \bibinfo{volume}{9596}, \bibinfo{publisher}{Springer International
  Publishing}, pp. \bibinfo{pages}{186--201},
  \doi{10.1007/978-3-319-31008-4_13}.

\bibitemdeclare{book}{HoFL}
\bibitem{HoFL}
\bibinfo{author}{G.~\surnamestart Rozenberg\surnameend} \&
  \bibinfo{author}{A.~\surnamestart Salomaa\surnameend} (\bibinfo{year}{1997}):
  \emph{\bibinfo{title}{Handbook of Formal Languages, Vol. 1: Word, Language,
  Grammar}}.
\newblock \bibinfo{publisher}{Springer-Verlag}, \bibinfo{address}{New York},
  \doi{10.1007/978-3-642-59126-6}.

\bibitemdeclare{book}{FL}
\bibitem{FL}
\bibinfo{author}{A.~\surnamestart Salomaa\surnameend} (\bibinfo{year}{1973}):
  \emph{\bibinfo{title}{Formal Languages}}.
\newblock \bibinfo{publisher}{Academic Press, London}.

\bibitemdeclare{inproceedings}{worth2005growing}
\bibitem{worth2005growing}
\bibinfo{author}{Peter \surnamestart Worth\surnameend} \&
  \bibinfo{author}{Susan \surnamestart Stepney\surnameend}
  (\bibinfo{year}{2005}): \emph{\bibinfo{title}{Growing Music: Musical
  Interpretations of L-Systems}}.
\newblock In \bibinfo{editor}{Franz \surnamestart Rothlauf\surnameend} et~al.,
  editors: {\slshape \bibinfo{booktitle}{Applications of Evolutionary
  Computing}}, {\slshape \bibinfo{series}{Lecture Notes in Computer Science}}
  \bibinfo{volume}{3449}, \bibinfo{publisher}{Springer}, pp.
  \bibinfo{pages}{545--550}, \doi{10.1007/978-3-540-32003-6_56}.

\bibitemdeclare{incollection}{zuidema2018formal}
\bibitem{zuidema2018formal}
\bibinfo{author}{Willem \surnamestart Zuidema\surnameend},
  \bibinfo{author}{Dieuwke \surnamestart Hupkes\surnameend},
  \bibinfo{author}{Geraint~A. \surnamestart Wiggins\surnameend},
  \bibinfo{author}{Constance \surnamestart Scharff\surnameend} \&
  \bibinfo{author}{Martin \surnamestart Rohrmeier\surnameend}
  (\bibinfo{year}{2018}): \emph{\bibinfo{title}{Formal Models of Structure
  Building in Music, Language, and Animal Song}}.
\newblock In \bibinfo{editor}{Henkjan \surnamestart Honing\surnameend}, editor:
  {\slshape \bibinfo{booktitle}{The Origins of Musicality}},
  \bibinfo{publisher}{MIT Press}, pp. \bibinfo{pages}{253--286},
  \doi{10.48550/arXiv.1901.05180}.

\end{thebibliography}

\end{document}